\documentclass[aps,prl,reprint,showpacs,groupedaddress,floatfix]{revtex4-1}
\usepackage{graphicx}
\usepackage{color}
\usepackage{hyperref}
\usepackage{amsmath,amssymb}
\usepackage[caption=false]{subfig}

\newcommand{\dd}{\mathrm{d}}
\newcommand{\du}[1]{\dd#1}
\providecommand*{\deriv}[3][]{
	\frac{\dd^{#1}#2}{\dd#3^{#1}}}
\providecommand*{\pderiv}[3][]{
	\frac{\partial^{#1}#2}{\partial#3^{#1}}}

\newcommand{\unit}[1]{\mathrm{#1}}
\newcommand{\bvec}[1]{\boldsymbol{\mathrm{#1}}}
\newcommand{\buvec}[1]{\hat{\bvec{#1}}} 
\newcommand{\eu}{\ensuremath{\mathrm{e}}}
\newcommand{\iu}{\ensuremath{i}}

\renewcommand\Im{\operatorname{\mathfrak{Im}}}

\newcommand{\gTensorXY}{\widehat{\mathcal{G}}_{xy}}
\newcommand{\oH}{\omega_H}
\newcommand{\oM}{\omega_M}
\newcommand{\dynMQXY}{\bvec{m}_{Q,xy}}
\newcommand{\gMix}{g_\perp}
\newcommand{\Heff}{\bvec{H}_\text{eff}}

\begin{document}

\title{Spin-pumping and Enhanced Gilbert Damping in Thin Magnetic Insulator Films}

\author{Andr\'{e} Kapelrud}
\author{Arne Brataas}
\affiliation{Department of Physics, Norwegian University of Science and Technology,
NO-7491 Trondheim, Norway}

\pacs{76.50.+g, 75.30.Ds, 75.70.-i, 75.76.+j, 75.78.-n}

\begin{abstract}
Precessing magnetization in a thin film magnetic insulator pumps spins into adjacent
metals; however, this phenomenon is not quantitatively understood. We present a
theory for the dependence of spin-pumping on the transverse mode number and in-plane
wave vector. For long-wavelength spin waves, the enhanced Gilbert damping for the
transverse mode volume waves is twice that of the macrospin mode, and for surface
modes, the enhancement can be ten or more times stronger. Spin-pumping is negligible
for short-wavelength exchange spin waves. We corroborate our analytical theory with
numerical calculations in agreement with recent experimental results.
\end{abstract}

\maketitle
Metallic spintronics have been tremendously successful in creating devices that both
fulfill significant market needs and challenge our understanding of spin transport in
materials. Topics that are currently of great interest are spin transfer and
spin-pumping \cite{Brataas:NatMat2012,PhysRevLett.88.117601, RevModPhys.77.1375},
spin Hall effects \cite{Jungwirth:NatMat2012}, and combinations thereof for use in
non-volatile memory, oscillator circuits, and spin wave logic devices.  A recent
experimental demonstration that spin transfer and spin-pumping can be as effective in
magnetic insulators as in metallic ferromagnetic systems was surprising and has
initiated a new field of inquiry \cite{Nature.464.262}. 

In magnetic insulators, no moving charges are present, and in some cases, the
dissipative losses associated with the magnetization dynamics are exceptionally low.
Nevertheless, when a magnetic insulator is placed in contact with a normal metal,
magnetization dynamics induce spin-pumping, which in turn causes angular momentum to
be dumped to the metal's itinerant electron system. Due to this non-local
interaction, the magnetization losses become enhanced. Careful experimental
investigations of spin-pumping and the associated enhanced magnetization dissipation
were recently performed, demonstrating that the dynamic coupling between the
magnetization dynamics in magnetic insulators and spin currents in adjacent normal
metals is strong. Importantly, in magnetic insulators, an exceptionally low intrinsic
damping combined with good material control has enabled the study of spin-pumping for
a much larger range of wave vectors than has previously been obtained in metallic
ferromagnets \cite{Nature.464.262, PhysRevLett.107.066604, burrowes:092403,
APL.97.252504, PhysRevLett.106.216601,
Rezende:apl2011,rezende:012402,Jia:EPL2011,2013arXiv1302.6697J, 2013arXiv1302.7091Q}.

In thin film ferromagnets, the magnetization dynamics are strongly affected by the
long-range dipolar interaction, which has both static and spatiotemporal
contributions. This yields different types of spin waves. When the in-plane
wavelength is comparable to the film thickness or greater, the long-range dipolar
interaction causes the separation of the spin-wave modes into three classes depending
on the relative orientation of the applied external field, in relation to the film
normal, and the spin-wave propagation direction \cite{PhysRev.118.1208, Damon1961308,
Puszkarski.IEEE.1973, wames:987, Kalinikos1986, Serga2010.43.264002}. These spin
waves are classified according to their dispersion and transverse magnetization
distribution as forward volume magnetostatic spin waves (FVMSWs) when the external
field is out-of-plane, backward volume magnetostatic spin waves (BVMSWs) when the
external field is in-plane and along the direction of propagation, and magnetostatic
surface waves (MSSWs) when the external field is in-plane but perpendicular to the
direction of propagation. In volume waves, the magnetic excitation is spatially
distributed across the entire film, while surface modes are localized near one of the
surfaces. ``Backward'' waves have a frequency dispersion with a negative group
velocity for some wavelengths. While these spin waves have been studied in great
detail over the last decades, the effect of an adjacent normal metal on these waves
has only recently been investigated.

Experimentally, it has been observed that spin-pumping differs for FVMSWs, BVMSWs and
MSSWs and that it depends on the spin-wave wavelength\cite{PhysRevLett.107.066604,
APL.97.252504, PhysRevLett.106.216601, Jia:EPL2011,2013arXiv1302.6697J,
2013arXiv1302.7091Q}. Recent experiments \cite{APL.97.252504} have demonstrated that
the magnetization dissipation is larger for surface spin waves in which the
excitation amplitude is localized at the magnetic insulator-normal metal interface.
To utilize spin-pumping from thin film magnetic insulators into adjacent normal
metals, a coherent theoretical picture of these experimental findings must be
developed and understood, which is the aim of our work.

\begin{figure}[bthp]
\centering
	\subfloat[][]{
		\includegraphics{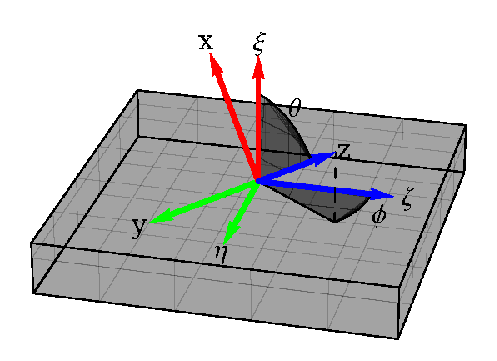} \label{fig:angles}
	}
	\subfloat[][]{
		\includegraphics{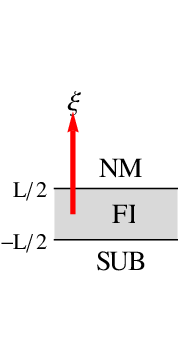} \label{fig:cross_section}
	}
	\caption{\label{fig:geometry}a) A thin film magnetic insulator of thickness $L$
	in its coordinate system; $\xi$ is the normal axis, the infinite
	$\eta\zeta$-plane is coplanar with the interfaces, and the spin waves propagate
	along the $\zeta$-axis. The internal field and saturation magnetization are along
	the $z$-axis. The $y$-axis is always kept in-plane, and the $x$-axis is selected
	such that the $x$-, $y$- and $z$-axes form a right-handed coordinate system. b)
	A cross-section showing the material stack.}
\end{figure}

In this Letter, we present a theory for energy dissipation from spin-wave excitations in a
ferromagnetic insulator (FI) thin film via spin-pumping when the ferromagnetic
insulator layer is in contact with a normal metal (NM). To this end, consider a thin
film magnetic insulator of thickness $L$ on an insulating substrate with a normal
metal capping (see Figure~\ref{fig:geometry}). We consider a normal metal such as Pt
at equilibrium, where there is rapid spin relaxation and no back-flow of spin
currents to the magnetic insulator. The normal metal is then a perfect spin sink and
remains in equilibrium even though spins are pumped into it.

The magnetization dynamics are described by the Landau-Lifshitz-Gilbert (LLG)
equation \cite{Gilbert1955} with a torque originating from the FI/NM interfacial
spin-pumping \footnote{Gaussian (cgs) units are employed throughout.}
\begin{equation}
	\dot{\bvec{M}} =
	-\gamma\bvec{M}\times\Heff
	+\frac{\alpha}{M_S}\bvec{M}\times\dot{\bvec{M}}
	+\bvec{\tau}_\text{sp},
	\label{eq:LLG}
\end{equation}
where $\alpha$ is the Gilbert damping coefficient, $M_S$ is the saturation
magnetization, $\gamma$ is the gyromagnetic ratio, $\Heff$ is the effective field
including the external field, exchange energy, surface anisotropy energy, and
static and dynamic demagnetization fields.

Spin-pumping through interfaces between magnetic insulators and normal metals
gives rise to a spin-pumping induced torque that is described as
\cite{PhysRevLett.88.117601}
\begin{equation}
	\bvec{\tau}_\text{sp}=\frac{\gamma\hbar^2}{2e^2M_S^2}\gMix\delta
		\Big(\xi-\frac{L}{2}\Big)\bvec{M}\times\dot{\bvec{M}},
	\label{eq:LLGp}
\end{equation}
where $\gMix$ is the transverse spin (``mixing'') conductance per unit area at the
FI/NM interface. We disregard the imaginary part of the mixing conductance because
this part has been found to be small at FI/NM interfaces \cite{Jia:EPL2011}. In
addition, the imaginary part is qualitatively less important and only renormalizes
the gyromagnetic ratio.

Assuming only uniform magnetic excitations, ``macrospin'' excitations, the effect of
spin-pumping on the magnetization dissipation is well known
\cite{PhysRevLett.88.117601, RevModPhys.77.1375}. Spin-pumping leads to enhanced
Gilbert damping, $\alpha\to\alpha+\Delta\alpha_\text{macro}$, which is proportional to
the FI/NM cross section because more spin current is then pumped out, but inversely
proportional to the volume of the ferromagnet that controls the total magnetic moment:
\begin{equation}
	\Delta\alpha_\text{macro}=\frac{\gamma\hbar}{4\pi LM_S}\frac{h}{e^2}\gMix.
	\label{eq:macrospinAlpha}
\end{equation}
Thus, the enhanced Gilbert damping due to spin-pumping is inversely proportional to
the film thickness $L$ and is important for thin film ferromagnets. However, a
``macrospin'' excitation, or the FMR mode, is only one out of many types of magnetic
excitations in thin films. The effect of spin-pumping on the other modes is not
known, and we provide the first analytical results for this important question, which
is further supported and complemented by numerical calculations. 

We consider weak magnetic excitations around a homogenous magnetic ground state
pointing along the direction of the internal field $\bvec{H}_i=H_i\buvec{z}$, which
is the combination of the external applied field and the static demagnetizing field
\cite{Kalinikos1986}. We may then expand
$\bvec{M}=M_S\buvec{z}+\bvec{m}_{Q,xy}(\xi)\eu^{\iu(\omega t-Q\zeta)}$, where
$\bvec{m}_{Q,xy}\cdot\buvec{z}=0$, $|\bvec{m}_{Q,xy}|\ll M_S$, and $Q$ is the
in-plane wave number in the $\zeta$-direction.
%

Following the linearization approach of the LLG equation \eqref{eq:LLG} as in
Ref.~\cite{Kalinikos1986}, we arrive at a two-dimensional integro-differential
equation of the dynamic magnetization (in the $xy$-plane) in the film's transverse
coordinate $\xi$:
\begin{widetext}
\begin{equation}
	\left[\iu\frac{\omega}{\oM}
	\begin{pmatrix}
		\alpha & -1 \\
		1 & \alpha
	\end{pmatrix}
	+\openone\left(
		\frac{\oH}{\oM}
			+8\pi\frac{\gamma^2A}{\oM^2}\left[Q^2-\deriv[2]{}{\xi}\right]
			+\iu\alpha\frac{\omega}{\oM}
	\right)\right]\dynMQXY(\xi) =
	\int_{-\frac{L}{2}}^{\frac{L}{2}}\du{\xi'}\,\gTensorXY(\xi-\xi')\dynMQXY(\xi'),
	\label{eq:motion}
\end{equation}
\end{widetext}
where $\omega$ is the spin-wave eigenfrequency, $A$ is the exchange stiffness, $\oH=\gamma H_i$,
$\oM=4\pi\gamma M_S$, and $\gTensorXY$ is the dipole-dipole field interaction tensor,
which fulfills the boundary conditions resulting from Maxwell's equations (see
\cite{Kalinikos1981}).

The eigensystem must be supplemented by boundary conditions that account for
spin-pumping and surface anisotropy. These boundary conditions are obtained by
integrating Eq.~\eqref{eq:LLG} over the interface \cite{RadoWeertman1959} and
expanding to the lowest order in the dynamic magnetization. When an out-of-plane easy
axis surface anisotropy is present, the boundary conditions are
\begin{subequations}
\label{eq:BClin}
\begin{gather}
	\left.\left(
		L\pderiv{}{\xi}+\iu\omega\chi+\frac{LK_s}{A}\cos(2\theta)
	\right) m_{Q,x}(\xi)\right|_{\xi=\frac{L}{2}}=0,
	\label{eq:BClinX} \\
	\left.\left(
		L\pderiv{}{\xi}+\iu\omega\chi+\frac{LK_s}{A}\cos^2(\theta)
	\right) m_{Q,y}(\xi)\right|_{\xi=\frac{L}{2}}=0,
	\label{eq:BClinY}
\end{gather}
\end{subequations}
where $K_s$ is the surface anisotropy energy with units $\unit{erg\cdot cm^{-2}}$ and
$\chi=\frac{L\hbar^2\gMix}{4Ae^2}$ is a parameter relating the exchange stiffness and
the spin mixing conductance ($[\chi]=\unit{s}$). The boundary condition at the
magnetic insulator--substrate interface at $\xi=-L/2$ is similar to
Eq.~\eqref{eq:BClin}, but simpler because $\chi\to0$ and $K_s\to0$ at that interface.

A mathematical challenge induced by spin-pumping arises because the second term in
the linearized boundary condition (\ref{eq:BClin}) is proportional to the eigenvalue
$\omega$ such that the eigenfunctions cannot simply be expanded in the set of
eigenfunctions obtained when there is no spin-pumping or dipolar interaction.
Instead, we follow an alternative analytical route for small and large wave vectors.
Furthermore, we numerically determine the eigenmodes with a custom-tailored
technique, where we discretize the differential equation (\ref{eq:motion}), include
the spin-pumping boundary conditions (\ref{eq:BClin}), and transform the resulting
equations into an eigenvalue problem in $\omega$ \cite{BrataasKapelrudUnpublished}.

Let us now outline how we obtain analytical results for small $QL\ll 1$ and large $QL
\gg 1$ wave vectors. First, we consider the case of vanishing surface anisotropy and
compute the renormalization of the Gilbert damping for the resulting modes. Next, we
demonstrate that the surface anisotropy creates a surface wave with a comparably
large enhancement of the Gilbert component.

When $QL\ll1$, the convolution integral on the right-hand side of
Eq.~\eqref{eq:motion} only contains the homogeneous demagnetization field. The
magnetization is then a transverse standing wave $\dynMQXY\left(\eu^{\iu
k\xi}+\eu^{-\iu k\xi+\phi}\right),$ where $k$ is a transverse wave number, $\phi$ is
a phase determined by the BC at the lower interface, and the two-dimensional
coefficient vector $\dynMQXY$ allows for elliptical polarization in the $xy$-plane.

By employing exchange-only boundary conditions \cite{RadoWeertman1959} at the lower
interface and using Eq.~\eqref{eq:BClin} with $K_s=0$ on the upper interface, the
transverse wave number $k$ is determined by $kL\tan{kL}=\iu\omega\chi$. Together with
the bulk dispersion relation $\omega =\omega(k)$, calculated from
Eq.~\eqref{eq:motion}, this expression allows us to calculate the magnetic excitation
dispersion relation parameterized by the film thickness, the Gilbert damping
$\alpha$, and the transverse conductance $\gMix$.

When spin-pumping is weak, $\omega\chi$ is small, and the solutions of the
transcendental equation can be expanded around the solutions obtained when there is
no spin-pumping, $kL=n\pi$, where $n$ is an integer. When $n\neq 0$, we expand to
first order in $kL$ and obtain $kL\approx n\pi+\iu\omega\chi/(n\pi)$.  When $n=0$, we
must perform a second-order expansion in terms of $kL$ around 0, which results in
$(kL)^2\approx \iu\omega\chi$. Using these relations in turn to eliminate $k$ from
the bulk dispersion relation while maintaining our linear approximation in small
terms and solving for $\omega$, we obtain complex eigenvalues, where the imaginary
part is proportional to a renormalized Gilbert damping parameter,
$\alpha^*=\alpha+\Delta\alpha$. When $n=0$, our results agree with the
spin-pumping-enhanced Gilbert damping of the macrospin (FMR) mode derived in
\cite{PhysRevLett.88.117601} (see Eq.~\eqref{eq:macrospinAlpha}),
$\Delta\alpha_0=\Delta \alpha_\text{macro}$. When $n\neq0$,  we compute
\begin{equation}
	\Delta\alpha_n=2\Delta\alpha_\text{macro}\,.
	\label{eq:deltaalpha}
\end{equation}
These new results indicate that {\it all} higher transverse volume modes have an
enhanced magnetization dissipation that is twice that of the macrospin mode. Thus,
counterintuitively, with the exception of the macrospin mode, increasingly
higher-order standing-wave transverse spin-wave modes have precisely the same
enhanced Gilbert damping.

Next, let us discuss spin-pumping for surface waves induced by the presence of
surface anisotropy.  When $K_s\neq0$, the lowest volume excitation mode develops into
a spatially localized surface wave. Expanding the expression for the localized wave
to the highest order in $L K_s/A$, we determine after some algebra that the resulting
enhancement of the Gilbert damping is
\begin{equation}
	\Delta\alpha_{n=0}=
		\frac{\gamma\hbar K_s}{4\pi M_sA}\frac{h}{e^2}\gMix
		\frac{\omega_H}{\omega_M}
		\left[
			\frac{\omega_H}{\omega_M}+\frac{1}{2}
			-\frac{K_s^2}{4\pi M_s^2A}
		\right]^{-1}.
	\label{eq:deltaalphaAnis}
\end{equation}
Comparing Eqs.~\eqref{eq:deltaalphaAnis} and \eqref{eq:deltaalpha}, we see that for
large surface anisotropy $L K_s/A \gg 1$, the spin-pumping-induced enhanced Gilbert
damping is independent of $L$. This result occurs because a large surface anisotropy
induces a surface wave with a decay length $A/K_s$, which replaces the actual
physical thickness $L$ as the effective thickness of the magnetic excitations, i.e.,
for surface waves $L\to A/K_s$ in the expression for the enhanced Gilbert damping of
Eq.\ (\ref{eq:macrospinAlpha}). This replacement implies that the enhanced Gilbert
damping is much larger for surface waves because the effective magnetic volume
decreases. For typical values of $A$ and $K_s$, we obtain an effective length
$A/K_s\sim 10\,\unit{nm}$. Compared with the film thicknesses used in recent
experiments, this value corresponds to a tenfold or greater increase in the
enhancement of the Gilbert damping. In contrast, for the volume modes ($n\neq0$), we
note from Eq.~\eqref{eq:BClin} that the dynamic magnetization will decrease at the
FI/NM interface due to the surface anisotropy; hence, $\Delta\alpha$ decreases
compared with the results of Eq.~\eqref{eq:deltaalpha}.

Finally, we can also demonstrate that for large wave vectors $QL\gg1$, the excitation
energy mostly arises from the in-plane (longitudinal) magnetization texture gradient.
Consequently, spin-pumping, which pumps energy out of the magnetic system due to the
transverse gradient of the magnetization texture, is much less effective and decays
as $1/(QL)^2$ with respect to Eq.~\eqref{eq:macrospinAlpha}.

To complement our analytical study, we numerically computed the eigenfrequencies
$\omega_n(Q)$. The energy is determined by the real part of $\omega_n(Q)$, while $\Im
\omega_n(Q)$ determines the dissipation rate and hence the spin-pumping contribution.
Recent experiments \cite{PhysRevLett.107.066604, 2013arXiv1302.6697J,
2013arXiv1302.7091Q, rezende:012402} on controlling and optimizing the ferrimagnetic
insulator yttrium-iron-garnet (YIG) have estimated that the mixing conductances of
both YIG|Au and YIG|Pt bilayers are in the range of $\gMix h/e^2\sim
0.02$--$3.43\cdot 10^{15}\,\unit{cm^{-2}}$. We use $\gMix h/e^2 = 1.2\cdot10^{14}
\,\unit{cm^{-2}}$ from Ref.~\cite{PhysRevLett.107.066604} in this work. All of our
results can be linearly re-scaled with other values of the mixing conductance. In the
following section, we also use $A=2.9\cdot10^{-8}\,\unit{erg/cm}$,
$K_s=0.05\,\unit{erg/cm^2}$, $L=100\,\unit{nm}$, $4\pi M_S=1750\,\unit{G}$, and
$\alpha=3\cdot10^{-4}$.

To distinguish the spin-pumping contribution $\Delta\alpha$ from  the magnetization
dissipation due to intrinsic Gilbert damping $\alpha$, we first compute the
eigenvalues, $\omega_\text{d}$, with intrinsic Gilbert damping, $\alpha\neq 0$, and
no spin-pumping, $\gMix=0$. Second, we compute the eigenvalues $\omega_\text{sp}$
with dissipation arising from spin-pumping only, $\alpha=0$ and $\gMix\neq0$. Because
$\Im\omega_\text{d}\propto\alpha$, we define a measure of the spin-pumping-induced
effective Gilbert damping as
$\Delta\alpha=\alpha\Im\omega_\text{sp}/\Im\omega_\text{d}$.

\begin{figure}
	\includegraphics[width=\linewidth]{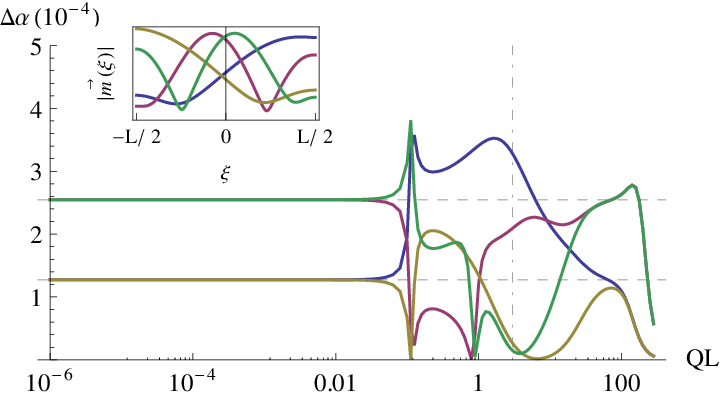}
	\caption{\label{fig:deltaAlphaMSSW}$\Delta\alpha$ versus wave vector for the MSSW
	geometry ($\theta=\phi=\pi/2$) for the four lowest eigenvalues.	Inset: Magnitudes
	of eigenvectors (in arbitrary units) across the film at $QL=1.5$.}
\end{figure}
We first consider the case of no surface anisotropy. Figure~\ref{fig:deltaAlphaMSSW}
shows the spin-pumping-enhanced Gilbert damping $\Delta\alpha$ as a function of the
product of the in-plane wave vector and the film thickness $QL$ in the MSSW geometry.
In the long-wavelength limit, $QL\ll1$, the numerical result agrees precisely with
our analytical results of Eq.~\eqref{eq:deltaalpha}. The enhanced Gilbert damping of
all higher transverse modes is exactly twice that of the macrospin mode. In the
dipole-exchange regime, for intermediate values of $QL$, the dipolar interaction
causes a small asymmetry in the eigenvectors for positive and negative
eigenfrequencies because modes traveling in opposite directions have different
magnitudes of precession near the FI/NM interface \cite{PhysRevLett.107.146602}, and
spin-pumping from these modes therefore differ. This phenomenon also explains why the
enhanced damping, $\Delta\alpha$, splits into different branches in this regime, as
shown in Fig.~\ref{fig:deltaAlphaMSSW}. For exchange spin waves, $QL\gg1$, the
exchange interaction dominates the dipolar interaction and removes mode asymmetries.
We also see that $\Delta\alpha\to0$ for large $QL$, in accordance with our analytical
theory. 

\begin{figure}
	\includegraphics[width=\linewidth]{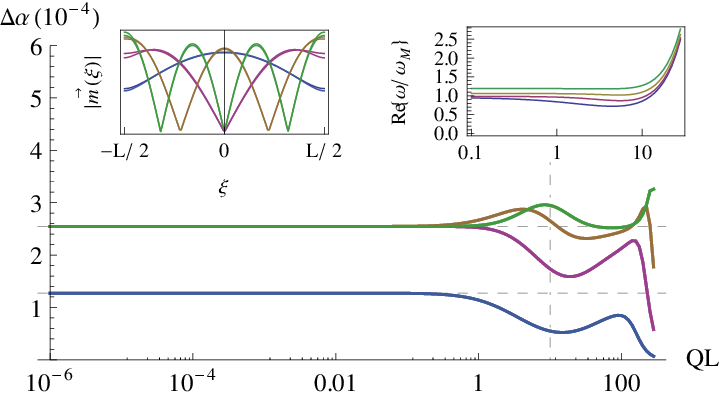}
	\caption{\label{fig:deltaAlphaBVMSW}$\Delta\alpha$ versus wave vector for the
	BVMSW geometry ($\theta=\pi/2$ and $\phi=0$). Left inset: Magnitude of
	eigenvectors (in arbitrary units) across the film when $QL=1.5$. Right inset: The
	real part of the dispersion relation for the same modes.}
\end{figure}
Figure~\ref{fig:deltaAlphaBVMSW} shows $\Delta\alpha$ for the BVMSW geometry. The
eight first modes are presented; however, as no substantial asymmetry exists between
eigenmodes traveling in different directions, the modes have the same pairwise
renormalization of $\alpha$. This symmetry occurs because the direction of the
internal field coincides with the direction of propagation. As in the previous case,
the dipolar interaction causes a slight shift in the eigenvectors in the intermediate
$QL$ regime, thereby altering $\Delta\alpha$ from that of Eq.~\eqref{eq:deltaalpha}.

\begin{figure}
	\includegraphics[width=\linewidth]{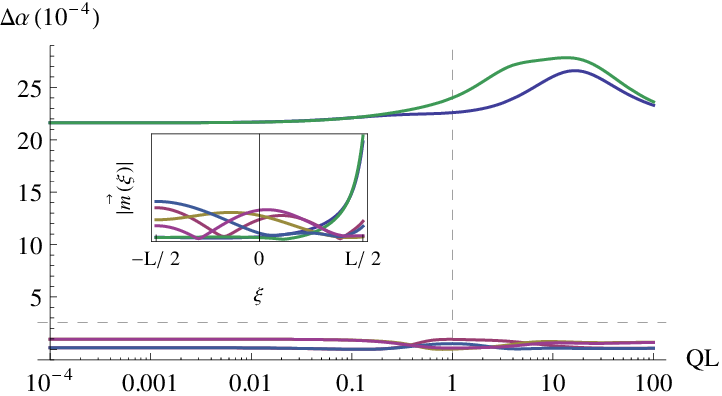}
	\caption{\label{fig:deltaAlphaMSSW_surf}$\Delta\alpha$ versus wave vector for 
	the MSSW geometry ($\theta=\phi=\pi/2$) with surface anisotropy added at the
	interface. Inset: Magnitudes of eigenvectors (in arbitrary units) across the
	film.}
\end{figure}
Figure~\ref{fig:deltaAlphaMSSW_surf} shows $\Delta\alpha$ for the MSSW geometry but
with surface anisotropy at the FI/NM interface. As expected from our analytical
results, surface anisotropy induces two localized surface modes with a ten-fold
larger enhancement of $\Delta\alpha$ compared with the volume modes. The horizontal
dashed line in Figure~\ref{fig:deltaAlphaMSSW_surf} indicates the analytical result
for the enhanced Gilbert damping of the $n\neq0$ modes when $K_s=0$. For the volume
modes, it is clear that the eigenvectors have a lower magnitude closer to the FI/NM
interface and that $\Delta\alpha$ is lower compared with the case of $K_s=0$, which
is consistent with our analytical analysis.

Our results also agree with recent experiments. \citet{APL.97.252504} found that
spin-pumping is significantly higher for surface spin waves compared with volume
spin-wave modes. In addition, in Ref.~\cite{PhysRevLett.106.216601}, exchange waves
were observed to be less efficient at pumping spins than dipolar spin waves, which is
consistent with our results. Furthermore, our results are consistent with the
theoretical finding that spin-transfer torques preferentially excite surface spin
waves with a critical current inversely proportional to the penetration depth
\cite{PhysRevLett.108.217204}.

In conclusion, we have analyzed how spin-pumping causes a wave-vector-dependent
enhancement of the Gilbert damping in thin magnetic insulators in contact with normal
metals. In the long-wavelength limit, our analytical results demonstrate that the
enhancement of the Gilbert damping for all higher-order volumetric modes is twice as
large as that of a macrospin excitation. Importantly, surface anisotropy-pinned modes
have a Gilbert renormalization that is significantly and linearly enhanced by the
ratio $L K_s/A$.
\begin{acknowledgments}
A. Kapelrud would like to thank G.\ E.\ W.\ Bauer for his hospitality at TU Delft.
This work was supported by EU-ICT-7 contract No. 257159 ``MACALO''.
\end{acknowledgments}


\clearpage
\end{document}